\documentclass[aps, pre, reprint, superscriptaddress, longbibliography]{revtex4-2}
\usepackage{graphicx}
\usepackage{amsmath}
\usepackage{amssymb}
\usepackage{dcolumn}
\usepackage{color}
\usepackage{bm}
\usepackage{braket}
\usepackage{txfonts}
\usepackage{xcolor}
\usepackage{hyperref}

\hypersetup{
	colorlinks=true,
	citecolor=blue,
	linkcolor=blue,
	urlcolor=blue,
}

\begin{document}
\title{Unsupervised classification of disordered patterns in an oppositely charged colloidal system}
\author{Yoshitaka Miyahara}
\email[]{sb22074r@st.omu.ac.jp}
\author{Taiki Haga}
\email[]{taiki.haga@omu.ac.jp}
\affiliation{Department of Physics and Electronics, Osaka Metropolitan University, Sakai-shi, Osaka 599-8531, Japan}
\date{\today}

\begin{abstract}
We develop an unsupervised machine learning approach to classify disordered patterns in a system of oppositely charged colloids. In this system, the interplay between Coulomb and van der Waals interactions leads to transitions in local structures, while the global structure remains disordered. Our method involves representing the local structures of the system as high-dimensional vectors and applying principal component analysis to identify distinct features of each pattern. We demonstrate that our method results in a reasonable classification of disordered patterns, which is consistent with that obtained from radial distribution functions. The interpretability of the method reveals the key characteristics of each pattern and provides valuable insights into the mechanisms underlying the unconventional phase transitions between disordered patterns.
\end{abstract}
\maketitle

\section{Introduction}

In recent years, machine learning is expected to become a powerful tool for the study of phase transitions.
To analyze phase transitions, it is essential to examine the statistical behavior of systems with many degrees of freedom. 
Machine learning is considered valuable for this purpose, as it excels in recognizing patterns within high-dimensional data.
In fact, numerous studies have demonstrated that machine learning can effectively identify phases across a variety of models.
These studies are divided into two categories based on the type of machine learning employed.
One uses supervised learning methods, while the other employs unsupervised learning methods. 
Supervised learning involves training on labeled data to classify new data based on these labels.
Therefore, in studies of phase transitions, supervised learning methods are effective when phase-labeled data is available.
For example, in previous studies on the Ising model and its variants, a fully connected neural network is trained to label spin configurations at high and low temperatures \cite{Carrasquilla2017}.
This approach enables accurate classification of data near the transition and precise determination of the transition temperature.
Similar methods have also been applied successfully to detect phases in quantum phase transitions and topological phase transitions \cite{Iakovlev2018,Venderley2018,Dong2019,Zhang2019}.

On the other hand, unsupervised learning does not require labeled data based on prior knowledge during training to identify underlying patterns in the data.
Thus, unsupervised learning is applicable even to systems exhibiting subtle and elusive changes in collective behavior, including not only phase transitions but also crossovers.
So far, especially for the Ising model, it has been shown that various unsupervised learning techniques can determine transition temperatures and order parameters based solely on spin configurations \cite{Torlai2016,Wang2016,Wetzel2017,Hu2017,Funai2020}. 
Furthermore, several studies have revealed that topological phase transitions can be detected by dimensional reduction techniques such as diffusion maps and variational autoencoders \cite{Rodriguez2019,Mathias2020,Che2020,Kaming2021,Mendes2021,Boattini2020}. 
Most previous studies have dealt with systems in which phases can be identified by known order parameters and topological invariants, and the features obtained by unsupervised learning have been interpreted based on prior knowledge of these quantities. 
However, in real-world applications, systems should be analyzed without any predefined order parameters, by relying solely on features extracted from unsupervised methods.
Yet, few studies have systematically addressed this scenario.
To resolve this gap, we apply unsupervised learning to a system exhibiting complex changes in collective behavior with no defined order parameter to demonstrate its practical applicability.

For our analysis, we focus on the dynamic pattern formation in colloidal systems.
These systems exhibit a wide range of spatio-temporal patterns, including oscillatory behaviors, hierarchical morphologies, and amorphous structures \cite{Saha2014, Mann2009, Bremer1993, Chen2011}.
Understanding how such out-of-equilibrium patterns emerge not only provides insight into fundamental physics but also facilitates the design of materials with novel functionalities \cite{Zhang2017, Avinshek2023}.
However, their complexity often makes it difficult for humans to extract meaningful descriptors that characterize the collective behavior underlying each pattern.
Therefore, analyzing colloidal patterns in nonequilibrium phases provides a valuable platform for evaluating the effectiveness of unsupervised learning in materials science applications.

Specifically, we consider a two-component mixture of oppositely charged colloids in two dimensions.
Simulations have shown that charged colloidal systems can exhibit various structures even under equilibrium conditions \cite{Imperio2006, Liu2008, Liu2019}, and recent advances in nanotechnology have made it easier to experimentally manipulate different types of colloids \cite{Perro2005,Sperling2010}.
When prepared from random initial conditions, the two-component colloidal systems often become trapped in long-lived metastable states due to high energy barriers that separate distinct global configurations.
In these nonequilibrium phases, although the overall structure remains disordered, distinct local structures emerge depending on the interactions.
The first objective of this study is to classify these structures by features derived from principal component analysis (PCA), a type of unsupervised learning.
We will investigate its effectiveness in identifying transitions between disordered phases. 
The second objective is to investigate what features of the phases are captured by PCA for classification.
By doing so, we can obtain insights into the properties that characterize the phases and the mechanisms underlying complex pattern changes.

In this study, we develop a method to transform configuration data, sampled at various parameter values through Monte Carlo simulations, into high-dimensional vectors representing the local structure around each colloid. 
We then apply PCA to these vectors, enabling us to identify local structures within the system.
Our analysis reveals that features extracted by PCA exhibit a nonmonotonic relationship with the interaction parameters.
The points where the trends shift between increasing and decreasing coincide with transition points independently identified using radial distribution functions.
Further examination reveals that PCA distinguishes the phases by highlighting structures in both average- and high-density regions.
We observe that transitions in local structure occur sequentially from low- to high-density regions within the system.

This paper is organized as follows. In Sec.~\ref{sec:model}, we provide a detailed description of our model.
Sec.~\ref{sec:phase_morphology} is dedicated to our examination of phases in our model and identifying them with radial distribution functions. 
In Sec.~\ref{sec:method}, we explain our proposed method utilizing unsupervised learning.
In Sec.~\ref{sec:result}, we present the results obtained by applying our method to the model and discuss our findings.
Finally, Sec.~\ref{sec:conclusion} provides conclusions and perspectives on future research.

\section{Model}
\label{sec:model}
We study a quasi two-dimensional (2D) system of oppositely charged colloids confined to a thin layer.
To simplify our simulations, we neglect the confined axis and model the system as strictly 2D, assuming that particle motion along that axis is negligible. 
This approximation is commonly used when the confinement is strong, as the in-plane coordinates of the particles dominate the structural and dynamical properties of the system \cite{Mejia2002,Tian2022}.
Thus, we perform strictly 2D simulations while employing an effective interaction potential for the quasi-2D geometry, based on DLVO theory.

The interactions between colloids charged with the same sign in an electrolyte solution have been understood based on DLVO theory, which includes the screened Coulomb and van der Waals interactions \cite{Ninham1999}.
Recent studies have shown that DLVO theory can also successfully describe interactions between oppositely charged colloids \cite{Cao2017,Cao2020,Sugimoto2022}.
A key characteristic of our system is that van der Waals forces between positively charged colloids are stronger than those in other pairs, causing them to attract each other more intensely.
As a result, the system shows various structures depending on the parameters of the Coulomb interactions.
In this paper, we apply unsupervised learning to quantitatively analyze these structural changes and contribute to a deeper understanding of systems with different types of colloids, which are widespread in both natural environments and industrial applications \cite{ALSHARIF2021}.

Our model considers two types of colloids P and N with a diameter $\sigma$.
A colloid P (N) is positively (negatively) charged and both are made from different materials.
The screened Coulomb interaction between two colloids at center-to-center distance $r_{ij}$ is given by
\begin{equation}\label{u_el}
    U_{\text{el}}(r_{ij})=\pi{}\epsilon{}\sigma\phi_i\phi_j\left(1+\frac{2\lambda}{\sigma}\right)^{2}\exp\left(-\frac{r_{ij}-\sigma}{\lambda}\right),
\end{equation}
where $\epsilon$ represents the solvent permittivity and $\phi_i$ is the electrostatic surface potential of the $i$th colloid which is represented by
\begin{equation}
    \phi_i=
    \begin{cases}
        \phi>0 & (\text{the $i$th colloid is a colloid P}), \\
        -\phi<0 & (\text{otherwise}),
    \end{cases}
\end{equation}
and $\lambda$ is the Debye screening length depending on the concentration of the electrolyte.
In Appendix \ref{sec:appendix_A}, we present the detailed derivation of Eq.~\eqref{u_el}.
In this study, we vary $\phi$ and $\lambda$, and investigate changes in patterns formed by colloids.

The van der Waals interaction $U_{\text{vdW}}(r_{ij})$ can be expressed as 
\begin{equation}\label{u_vdW}
    U_{\text{vdW}}(r_{ij})=-\frac{A_{ij}\sigma}{24(r_{ij}-\sigma)},
\end{equation}
where $A_{ij}$ is referred to as the Hamaker constant and it depends on the types of interacting colloids. 
We can calculate $A_{ij}$ approximately via 
\begin{equation}
    A_{ij}=\left(\sqrt{A_{i}}-\sqrt{A_{\text{sol}}}\right)\left(\sqrt{A_{j}}-\sqrt{A_{\text{sol}}}\right),
\end{equation}
where $A_{i}$ and $A_{\text{sol}}$ represent the strength of the van der Waals interactions that act between molecules constituting the $i$th colloid and the solvent, respectively, in a vacuum \cite{Israelachvili2011}.

We consider a situation in which the range of the Coulomb interaction, determined by $\lambda$, is shorter than the size of the colloids.
Since the van der Waals interaction is inherently short-ranged, the interaction potential $U(r_{ij})$ becomes negligible for $r_{ij} \ge 4\sigma$.
To prevent the divergence of the potential energy that could result in unphysical aggregation of colloids, a cutoff length of $1.01\sigma$ is introduced.
Considering them, the interaction potential is described as
\begin{equation}
    U(r_{ij})=
    \begin{cases}
        \infty & (r_{ij}\leq1.01\sigma), \\
        U_{\text{el}}(r_{ij})+U_{\text{vdW}}(r_{ij}) & (1.01\sigma < r_{ij}<4\sigma), \\
        0 & (r_{ij}\geq4\sigma).
    \end{cases}
\end{equation}
We have confirmed that extending the cutoff beyond $4\sigma$ does not alter the statistical properties of the system.

Under the condition of short interaction range, it is justified to use the same interaction form in our quasi-2D system as in 3D systems.
However, the parameters of the potential can be modified depending on the properties of the confining plates.
For example, a previous study has shown that colloids confined between highly charged glass plates experience stronger attractive interactions than in the 3D bulk \cite{Kepler1994}.
To account for this effect, one would need to use renormalized parameters.
Since our primary objective is to capture the qualitative behavior of oppositely charged colloids, we have adopted the same interaction parameters as those for 3D bulk systems.

We investigate energetically stable configurations of colloids at room temperature under given parameters using the Monte Carlo simulation \cite{Dickinson1992}.
At each time step in this simulation, a colloid is chosen randomly and displaced by $\Delta{}\boldsymbol{x}$, where $\Delta{}\boldsymbol{x}$ is sampled from a normal distribution with a mean of zero and a variance of $\sigma^2$.
The probability $P$ of accepting this move is determined by the energy change $\Delta{}E$ induced by the displacement, the Boltzmann constant $k_B$, and temperature $T$ as
\begin{equation}
  P=\min\left\{1,\exp{}\left(-\frac{\Delta{}E}{k_BT}\right)\right\}.
\end{equation}
The initial configuration is set up randomly, ensuring that the center-to-center distance of every pair of colloids is greater than $1.01\sigma$.
This single-particle Monte Carlo approach qualitatively reproduces the overdamped dynamics of colloids and is generally more stable than Langevin molecular dynamics simulations~\cite{Sanz2010}.  
Therefore, it serves as a suitable method for sampling dynamically frozen states that would be generated by specific operational protocols in real time.

The simulation is conducted in the system size $24\sigma\times24\sigma$, containing 105 colloids of each type, resulting in a total number $N_{\text{tot}} = 210$. 
The other parameters for the simulation are listed in Table \ref{tab:simulation_paras}.
The Hamaker constants for colloids are chosen to be close to those of metallic materials \cite{Israelachvili2011}.
\begin{table}[t]
  \centering
  \caption{List of parameters in our simulation}
  \label{tab:simulation_paras}
  \begin{tabular}{cc} \hline
    \begin{tabular}{l} 
      diameter $\sigma$
      \\(nm)
  \end{tabular}
  & 100
  \\ \hline
  \begin{tabular}{l} 
      solvent permittivity $\epsilon$
    \\($\times10^{-10}$ F/m)
  \end{tabular}
  & 6.94
  \\ \hline
  \begin{tabular}{l}
      surface potential $\phi$
    \\(mV)
  \end{tabular}
  & 1 \text{--} 20
  \\ \hline
  \begin{tabular}{l}
      debye length $\lambda$
    \\(nm)
    \end{tabular}
  & 2.5 \text{--} 50
  \\ \hline
  \begin{tabular}{l}
      Hamaker constant of positive colloids $A_p$
    \\($\times10^{-20}$~J)
  \end{tabular}
  & 18
  \\ \hline
  \begin{tabular}{l}
      $\hdots$ of negative colloids $A_n$
    \\($\times10^{-20}$~J)
  \end{tabular}
  & 10 
  \\ \hline
  \begin{tabular}{l}
    $\hdots$ of solvent $A_{\text{sol}}$
    \\($\times10^{-20}$~J)
  \end{tabular}
  & 4
  \\ \hline
  \begin{tabular}{l}
      temperature $T$
    \\(K)
  \end{tabular}
  & 298
  \\ \hline
\end{tabular}
\end{table}

In general, two-component colloidal systems like ours tend to struggle to form long-range crystalline order and instead exhibit locally favored patterns within a globally disordered configuration.
When starting our simulations from an initial random configuration, we observe two distinct relaxation processes.
The first occurs on a shorter timescale, during which colloids rapidly form locally favored patterns, leading to a dynamically frozen state.
Following this initial stage, the colloids relax toward equilibrium over a longer timescale.
In this study, we focus on the dynamically frozen patterns that emerge after the first relaxation process.
To extract the characteristic timescales and determine the appropriate sampling time, we fit the time evolution of the potential energy using a multi-exponential function.
In Appendix \ref{sec:appendix_B}, we present the detailed descriptions of the sampling procedures and the system dynamics around the sampling time.
We generate every 1500 samples, using different initial conditions, at $\phi = 1, 2, \cdots, 20$ mV and $\lambda = 2.5, 5, \cdots, 50$ nm.

\section{Phase morphology}
\label{sec:phase_morphology}

%%%%%%%%%%%%%%%%%%%%%%%%%%%%%%%%%%%%
\begin{figure*}
\centering
\includegraphics[width=\textwidth]{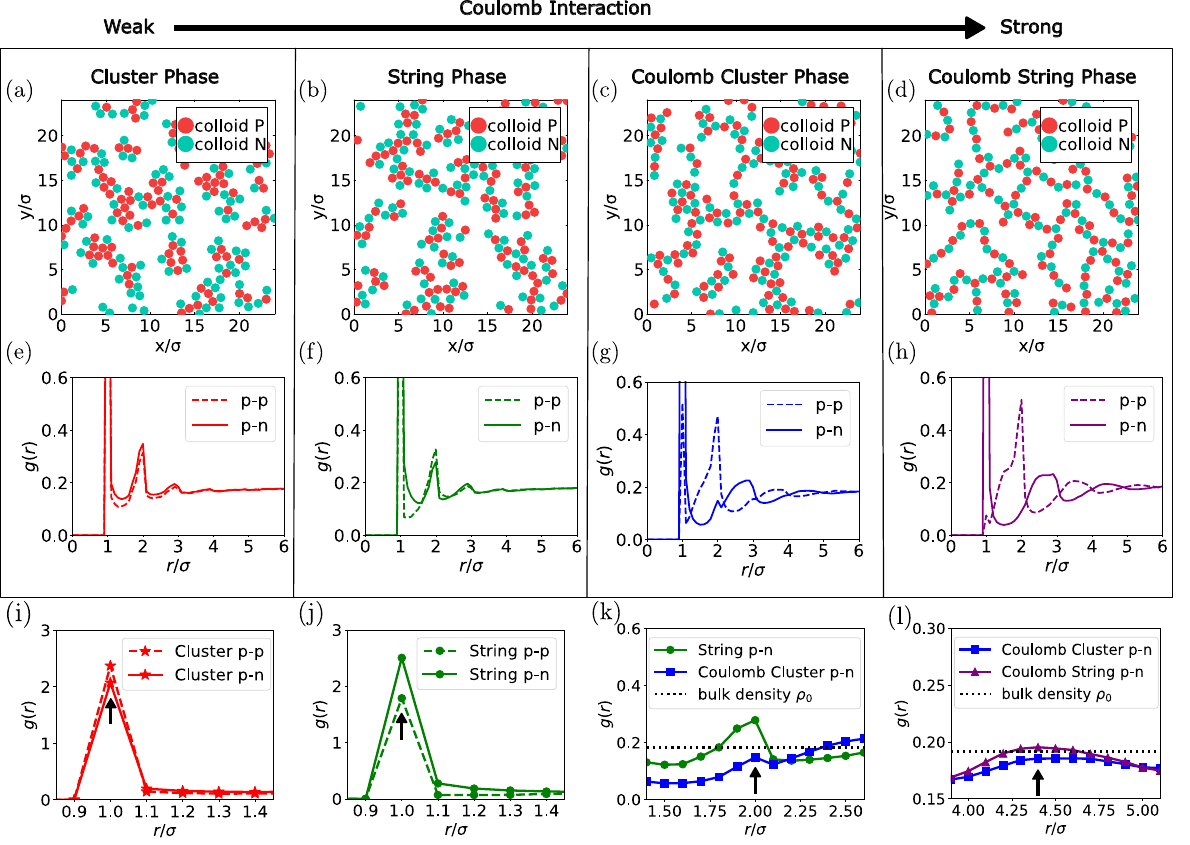}
\caption{Panels (a)-(d) show typical configurations at $\lambda = 40$~nm for the Cluster ($\phi=1$~mV), String ($\phi = 6$~mV), Coulomb Cluster ($\phi = 13$~mV), and Coulomb String phases ($\phi = 20$~mV), respectively. Red particles are positively charged colloids, namely colloids P, while Green ones are negatively charged colloids, colloids N. Panels (e)-(h) display the corresponding RDFs; $g_{pn}$ (solid line) and $g_{pp}$ (dashed line). Panels (i)-(l) highlight the differences in $g_{pn}$ (solid) and $g_{pp}$ (dashed) across different phases, with black arrows indicating reference points used to identify phase boundaries. Panels~(i) and (j) show that $g_{pn}(\sigma)$ becomes larger than $g_{pp}(\sigma)$ as the system transitions from the Cluster phase to the String phase. Panel~(k) illustrates that $g_{pn}(2\sigma)$ becomes smaller than the bulk density $\rho_0$ as the system transitions from the String phase to the Coulomb Cluster phase. Panel~(l) shows that $g_{pn}(r)$ in the Coulomb String phase exhibits a distinct peak at $r > 4\sigma$, which is not observed in the Coulomb Cluster phase.}
\label{fig:phase_morphology}
\end{figure*}
%%%%%%%%%%%%%%%%%%%%%%%%%%%%%%%%%%%%

In our model, we can observe four distinct nonequilibrium phases resulting from the competition between the Coulomb and van der Waals interactions.
For our purposes, we refer to a phase as a region of a many-particle system that is macroscopically uniform in its physical properties, regardless of whether its boundaries are marked by singular behavior in any physical quantities.
Furthermore, the nonequilibrium phases are characterized not only by the interaction parameters, but also by the specific sampling protocols, in contrast to equilibrium phases. 
Figures \ref{fig:phase_morphology}(a)-(d) show that as the parameter $\phi$ varies at fixed $\lambda$, the local patterns of colloids evolve as follows.
\begin{enumerate}
\item[(a)]Cluster phase: Colloids P aggregate into clusters.
\item[(b)]String phase: Colloids P form elongated, string-like structures.
\item[(c)]Coulomb Cluster phase: Clusters resembling ionic crystals disperse.
\item[(d)]Coulomb String phase: Colloids P and N are arranged alternately in string-like structures.
\end{enumerate}
These phases can be roughly distinguished by focusing on the local structures of colloids P. However, the local structures across the entire system are not uniform, making it difficult to precisely define boundaries between phases.
It is worth noting that in thermal equilibrium, the system exhibits phases significantly different from nonequilibrium ones shown in Fig.~\ref{fig:phase_morphology}.
In Appendix~\ref{sec:appendix_C}, we present equilibrium phases obtained by simulated annealing.

In the next section, we aim to identify the precise boundaries between the observed phases using an unsupervised learning method.
To provide a baseline for comparison, we first estimate the phase boundaries using traditional statistical mechanics quantities.
Specifically, we use the radial distribution function (RDF), which characterizes the local structure of colloids. 
The RDF $g(r)$ represents a colloidal density at a distance $r$ from a reference colloid and is defined as
\begin{equation}
    g(r)=\frac{dN(r)}{2\pi{r}dr},
\end{equation}
where $dN(r)$  is the number of colloids within a shell of thickness $dr$ at a distance $r$ from the reference colloid ($dr = 0.1\sigma$).
In this study, we consider two RDFs: $g_{pp}(r)$, the density of a colloid P around another colloid P, and $g_{pn}(r)$, the density of colloid N around a colloid P [see Fig.~\ref{fig:phase_morphology}(e)-(h)].
In the rest of this section, we describe the procedure for defining each phase boundary.

A straightforward approach to identify the boundary between the Cluster and String phases is to count the number of colloids P in contact with a given colloid P.
However, since the presence of colloids N suppresses the clustering of colloids P, no significant difference is observed in this contact number across phases.
Instead, we focus on the difference in contact ratios: in the String phase, fewer colloids P are in contact with a given colloid P compared to colloids N, whereas the opposite trend occurs in the Cluster phase.
Based on this observation, we define the phase boundary as the parameter value at which $g_{pn}(\sigma) / g_{pp}(\sigma) = 1$ [see Fig.~\ref{fig:phase_morphology}(i) and (j)].
Specifically, parameter regions where $g_{pn}(\sigma) / g_{pp}(\sigma) < 1$ are identified as the Cluster phase, while regions where $g_{pn}(\sigma) / g_{pp}(\sigma) \geq 1$ are classified as the String phase.

To identify the boundary between the String and Coulomb Cluster phases, we focus on the formation of ionic crystal-like clusters, which characterize the Coulomb Cluster phase.
As the strength of the Coulomb interaction increases from the String phase, the transition point is identified as the parameter value where clusters start to form.
In ionic crystals, ions of the same charge are arranged at intervals of $2\sigma$ along orthogonal directions.
We find that $g_{pn}(2\sigma)$ is lower than the bulk density $\rho_{0}$.
Considering that, we define the phase boundary as the point where $g_{pn}(2\sigma) = \rho_{0}$ [see Fig.~\ref{fig:phase_morphology}(k)].

In the Coulomb String phase, colloids P and N alternate in a string-like structure that extends over long distances.
This long-range order can be detected through the behavior of $g_{pn}(r)$ at large $r$.
As the Coulomb interaction strength increases from the Coulomb Cluster phase, a peak in $g_{pn}$ emerges at distances beyond $4\sigma$, indicating the onset of long-range order.
To account for noise, we introduce a threshold value $\rho_{\text{th}}=1.05\rho_{0}$ and the Coulomb String phase is then identified as a parameter region where there exists $r>4\sigma$ such that $g_{pn}(r)>\rho_{\text{th}}$ [see Fig.~\ref{fig:phase_morphology}(l)].
By comparing the transition points determined above with those obtained through our unsupervised learning method, we discuss its validity and effectiveness.

%%%%%%%%%%%%%%%%%%%%%%%%%%%%%%%%%%%%
\begin{figure*}
 \includegraphics[width=0.85\linewidth]{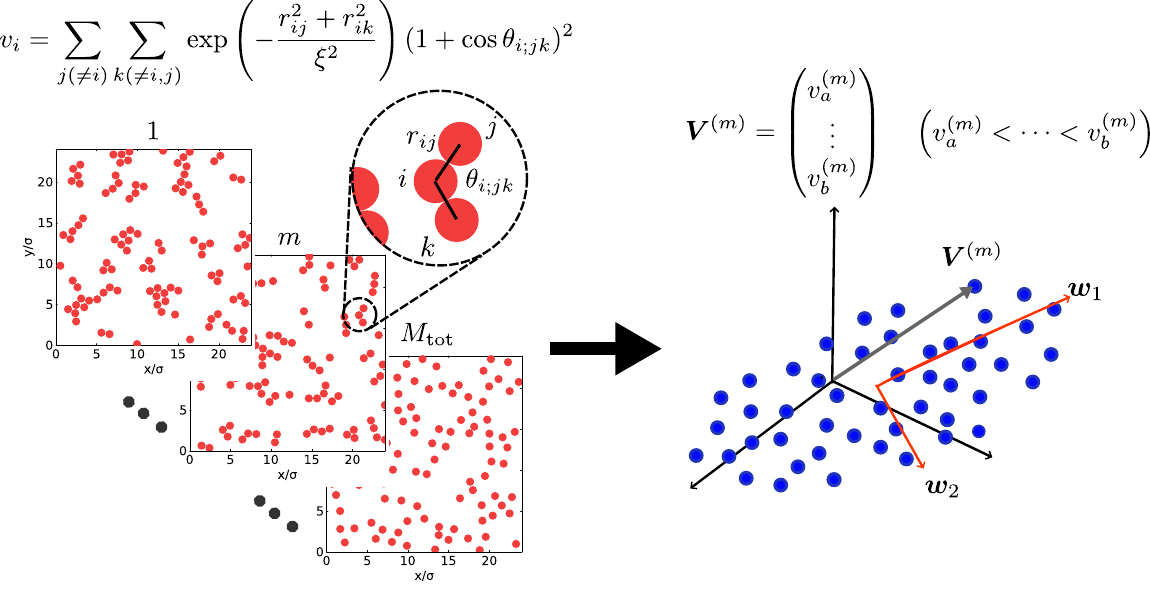}
  \caption{Schematic illustration of our method. First, configurations are generated for each set of parameters using the Monte Carlo method. Next, the descriptors $v$ are calculated based on the local structures formed by colloids P. These descriptors are sorted in ascending order for each configuration to form vectors $\boldsymbol{V}$. Finally, PCA is applied to these vectors.}
  \label{fig:method}
\end{figure*}
%%%%%%%%%%%%%%%%%%%%%%%%%%%%%%%%%%%%

\section{method}
\label{sec:method}

Our method involves representing the local structures of colloids as vectors and applying PCA to extract phase features.
As described in Sec.~\ref{sec:phase_morphology}, the phases are characterized by the local arrangements of colloids P, so we focus exclusively on the local structures formed by colloids P.
For the $i$th colloid P, we define the local structure descriptor $v_i$ as
\begin{equation}\label{V}
  v_i=\sum_{j(\ne{}i)}\sum_{k(\ne{}i,j)}\exp\left({-\frac{r_{ij}^2+r_{ik}^2}{\xi^2}}\right)(1+\cos\theta_{i;jk})^2,
\end{equation}
where $\theta_{i;jk}$ represents the angle formed by the $j$th and $k$th colloids P with respect to the $i$th colloid P, and $\xi = 2\sigma$ is the cutoff parameter \cite{Behler2011,Cubuk2015,Caro2018,Tamura2022}.
To reduce computational costs, the sums are restricted to $r_{ij}<10\sigma$ and $r_{ik}<10\sigma$.
The descriptor $v_i$ correlates with the number of colloids P surrounding the $i$th colloid P.

PCA identifies a set of orthogonal axes in data, where each axis represents a direction of large variance \cite{Pearson1901}.
The axis $\boldsymbol{w}_l$ with the $l$th largest variance is called the $l$th principal component axis.
For a dataset represented by an $n \times m$ matrix $X$, where $n$ is the number of samples and $m$ is the number of components, $\boldsymbol{w}_l$ are obtained by solving the eigenvalue equation:
\begin{equation}
(\boldsymbol{CX})^T\boldsymbol{CX}\boldsymbol{w}_l=\lambda_l\boldsymbol{w}_l,
\end{equation}
where $\lambda_l$ is the $l$th largest eigenvalue and $\boldsymbol{C}$ is the centering matrix, ensuring that each column of $\boldsymbol{CX}$ has a mean of zero.
$\boldsymbol{C}$ is defined as
\begin{equation}
  \boldsymbol{C}=\boldsymbol{I}_n-\frac{1}{n}\boldsymbol{1}_{n}\boldsymbol{1}_{n}^T,
\end{equation}
where $\boldsymbol{I}_n$ represents the $n\times{}n$ identity matrix and $\boldsymbol{1}_n$ is an $n$-dimensional column vector with all entries equal to 1.
The features extracted by PCA are expressed as the $l$th principal components (PCs) $\boldsymbol{y}_l=\boldsymbol{X}\boldsymbol{w}_l$, which represent the projection of each sample onto the $l$th principal component axis.
In this study, we focus on the 1st and 2nd PCs.

The detailed procedure of our method is outlined in the following steps [see Fig.~\ref{fig:method}]:
\begin{enumerate}
\item[1.]Sample $M$ configurations with different initial conditions for fixed $\lambda$ and $\phi$ using the Monte Carlo simulation.
\item[2.]Iterate step 1 varying $\lambda$ and $\phi$, and collect $M_{\text{tot}}$ configurations in total.
\item[3.]Calculate the descriptor $v_i^{(m)}$ by Eq.~\eqref{V} for each colloid P in the $m$th configuration.
\item[4.]Sort the descriptors of the $m$th configuration in ascending order, and obtain the vector $\boldsymbol{V}^{(m)}$, whose dimension is equal to the number $N_p$ of colloids P. 
\item[5.]Apply PCA to the vectors obtained by iterating steps 3 and 4.
\end{enumerate}
As described in Sec.~\ref{sec:model}, we set $M = 1500$, $\phi = 1, 2,\cdots, 20$~mV, $\lambda =2.5, 5, \cdots, 50$~nm and $N_p = 105$.
Therefore, the total number $M_{\text{tot}}$ of configurations is $1500\times20\times20$ and the dimension of $\boldsymbol{V}$ is 105.
In this study, we focus on the structure that remains invariant under translational and rotational transformations.
To ensure the vector $\boldsymbol{V}$ shares this invariance, we sort the descriptors in ascending order.
By applying PCA to these vectors, we can analyze the distribution of features that characterize the local structures.

By using machine learning techniques with local structural descriptors, recent studies have elucidated the relationships between local atomic structures and physical or chemical properties \cite{Cubuk2015, Caro2018, Boattini2020}.
In these studies, the local structure around a target atom is transformed into a vector to serve as input data for machine learning.
In contrast, our method represents the local structures of all particles in a snapshot as a single vector and applies PCA.
Since phase transitions arise from the collective behavior of all components in a system, our method is well-suited for phase classification.
However, as our approach yields only one data point from a single configuration, collecting a large dataset can be challenging.
With advancements in simulation acceleration techniques, we believe integrating our method with these technologies could significantly improve its efficiency and applicability \cite{Liu2017, Huang2017, Nagai2020}.

\section{result}
\label{sec:result}

\begin{figure}
 \includegraphics[width=\linewidth]{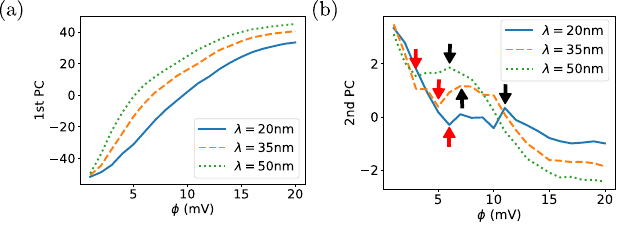}
  \caption{The 1st [panel (a)] and 2nd [panel (b)] PCs averaged for each set of parameters. In panel (b), red arrows indicate $\phi_{\text{c-s}}$, where the trend shifts from decreasing to increasing, while black ones mark $\phi_{\text{s-cc}}$, where the reverse shift occurs.}
  \label{fig:result1}
\end{figure}

Our method transforms each configuration into a two-dimensional vector comprising the 1st and 2nd PCs.
If this method effectively captures the characteristics of the patterns formed by colloids, the PC values should exhibit notable behavior near the parameters where pattern changes occur.
To investigate this, we average the PCs for each set of interaction parameters, $\phi$ and $\lambda$, and the results are shown in Fig.~$\ref{fig:result1}$.
Figures $\ref{fig:result1}$(a) and $\ref{fig:result1}$(b) display the 1st and 2nd PCs as functions of $\phi$ for different values of $\lambda$.
The 1st PC increases monotonically with $\phi$ and shows no singular behavior.
In contrast, the 2nd PC exhibits two distinct points, $\phi_{\text{c-s}}$ and $\phi_{\text{s-cc}}$, where the trends shift between increasing and decreasing.
At $\phi_{\text{c-s}}$, the trend of the 2nd PC transitions from decreasing to increasing, whereas at $\phi_{\text{s-cc}}$, it shifts from increasing to decreasing.
In fact, observations of the configurations suggest that $\phi_{\text{c-s}}$ corresponds to the transition point between the Coulomb and String phases, and $\phi_{\text{s-cc}}$ is close to the transition point between the String and Coulomb Cluster phases.
To verify this quantitatively, we compare $\phi_{\text{c-s}}$ and $\phi_{\text{s-cc}}$ with the transition points identified using RDFs.

%%%%%%%%%%%%%%%%%%%%%%%%%%%%%%%%%%%%
%%%%%%%%%%%%%%%%%%%%%%%%%%%%%%%%%%%%

\begin{figure}
 \includegraphics[width=\linewidth]{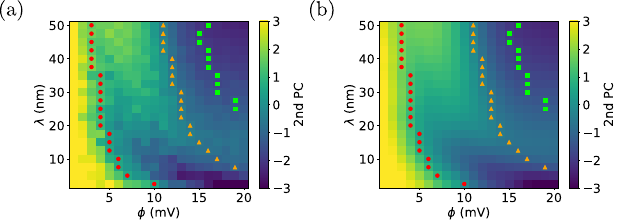}
  \caption{Heatmaps of the 2nd PC, with the vertical axis representing $\lambda$ and the horizontal axis representing $\phi$. Panel (b) shows the same data as panel (a), smoothed using a Gaussian filter. As $\phi$ increases at a fixed $\lambda$, the 2nd PC decreases until it reaches the valley, $\phi_{\text{c-s}}$. As $\phi$ continues to increase, the 2nd PC begins to increase and transitions to a decrease at the ridge, $\phi_{\text{s-cc}}$. The phase boundaries identified by RDFs are marked as red circles for the Cluster-String boundary, orange triangles for the String-Coulomb Cluster boundary, and green squares for the Coulomb Cluster-Coulomb String boundary.}
  \label{fig:result2}
\end{figure}

Figure \ref{fig:result2}(a) shows the 2nd PC as a function of $\phi$ and $\lambda$, with Fig.~\ref{fig:result2}(b) presenting the same data smoothed using a Gaussian filter.
As discussed above, the valley on the left side of the heatmap corresponds to $\phi_{\text{c-s}}$, while the central ridge corresponds to $\phi_{\text{s-cc}}$.
The phase transitions in our system are driven by the competition between Coulomb and van der Waals interactions.
Consequently, as $\lambda$ increases, the true transition points are expected to shift to smaller values of $\phi$.
The figures clearly show that the behaviors of $\phi_{\text{c-s}}$ and $\phi_{\text{s-cc}}$ along $\lambda$ are consistent with this expectation.

%%%%%%%%%%%%%%%%%%%%%%%%%%%%%%%%%%%%
%%%%%%%%%%%%%%%%%%%%%%%%%%%%%%%%%%%%

The symbols in Fig.~\ref{fig:result2} indicate the transition points determined by RDFs.
Firstly, the transition points between the Cluster and String phases, which are denoted by red circles, coincide well with those determined by the 2nd PC at larger $\lambda$.
While significant deviations are observed for smaller $\lambda$, the trends in the shifts of the phase boundaries with respect to $\lambda$ remain consistent.
Second, the transition points between the String and Coulomb Cluster phases, indicated by yellow triangles, also show qualitative agreement with those identified by the 2nd PC.
Furthermore, the transition points between the Coulomb Cluster and Coulomb String phases (green squares) determined by RDFs are found along the contour lines enclosing the black region in the upper-right corner of the heatmap.
These results suggest that the 2nd PC can effectively classify transitions between these phases.
In the remainder of this section, we discuss the causes of the trend shifts observed in the 2nd PC, which are associated with the Cluster-String and String–Coulomb Cluster phase boundaries.
Clarifying the relationship between the Coulomb Cluster–Coulomb String phase boundary and the 2nd PC remains an open question for future work.

\begin{figure}
 \includegraphics[width=\linewidth]{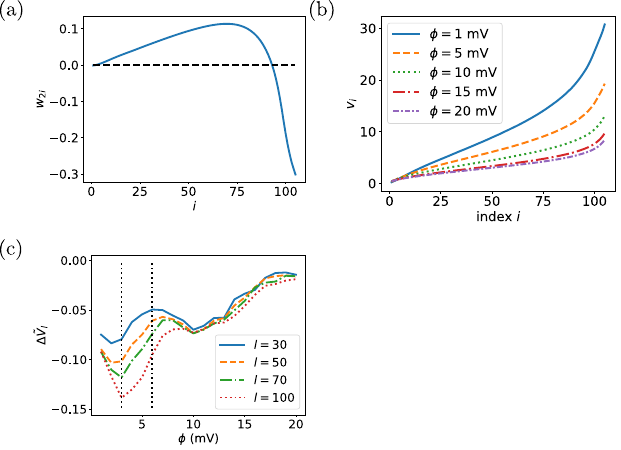}
  \caption{Panel (a) shows the weights $w_{2i}$ of the second principal axis $\boldsymbol{w}_2$. Panel (b) plots the averaged vector $\boldsymbol{V}$ for each $\phi$ at $\lambda = 50$~nm. Panel (c) presents the dependence of $\Delta\tilde{v}_i$ on $\phi$ at $i = 30, 50, 70, 100$. Two dotted lines indicate $\phi_{\text{c-s}}$ and $\phi_{\text{s-cc}}$ from left to right}
  \label{fig:result3}
\end{figure}

Since the 2nd PC is the inner product of the second principal axis $\boldsymbol{w}_2$ and the data vector $\boldsymbol{V}$, each component of $\boldsymbol{w}_2$ acts as a weight for the corresponding component of $\boldsymbol{V}$.
To identify which local structural descriptors affect the 2nd PC, we examine the weight $w_{2i}$, which is the $i$th component of $\boldsymbol{w}_2$.
Specifically, we focus on the regions near the first two transition points (excluding the transition between the Coulomb Cluster and Coulomb String phases), where the 2nd PC changes its trend.
Note that in $\boldsymbol{V}$, the descriptors are sorted in ascending order, and each component $v_i$ correlates with the local density around the target colloid.
Therefore, smaller indices $i$ correspond to low-density regions, while larger indices reflect high-density regions.
Figure \ref{fig:result3}(a) plots $w_{2i}$ as a function of the index $i$.
It can be observed that $w_{2i}$ exhibits relatively large positive values in the middle range ($i = 50 \text{--} 80$) and significant negative values in the higher range ($i = 95 \text{--} 105$).
(In the following, we refer to the representative element in the set of $v_i$ for $i = 50\text{--}80$ as $v_{med}$, and that for $i = 95\text{--}105$ as $v_{lar}$.)
This indicates that descriptors from low- and high-density regions contribute to the 2nd PC with opposite signs.
More specifically, the 2nd PC is approximately proportional to the difference between $v_{\text{med}}$ and $v_{\text{lar}}$, which represent $v_i$ corresponding to medium- and high-density regions.

%%%%%%%%%%%%%%%%%%%%%%%%%%%%%%%%%%%%
%%%%%%%%%%%%%%%%%%%%%%%%%%%%%%%%%%%%

To understand the non-monotonic behavior of the 2nd PC shown in Fig.~\ref{fig:result1}(b), we discuss how $v_{\text{med}}$ and $v_{\text{lar}}$ change with increasing $\phi$.
Figure \ref{fig:result3}(b) shows $v_i$ for various values of $\phi$. For all indices $i$, $v_i$ decreases with $\phi$ because stronger Coulomb interactions prevent positively charged colloids from clustering and reduce their local density.
Since $v_{\text{med}}$ and $v_{\text{lar}}$ contribute the 2nd PC with opposite signs, the 2nd PC increases with $\phi$ when $v_{\text{lar}}$ decreases more rapidly than $v_{\text{med}}$.
To confirm this expectation, we define the rate of change $\Delta \tilde{v}_i$ with respect to $\phi$ as:
\begin{equation}
  \Delta\tilde{v}_i(\phi)=
  \begin{cases}
\frac{v_i(\phi+\Delta\phi) - v_i(\phi)}{v_i(\phi)}  & (\phi=1\text{~mV}), \\
\frac{v_i(\phi) - v_i(\phi-\Delta\phi)}{v_i(\phi)} & (\phi=20\text{~mV}), \\
\frac{v_i(\phi+\Delta\phi) - v_i(\phi-\Delta\phi)}{2v_i(\phi)} & \text{(otherwise),} 
\end{cases}
\end{equation}
where $\Delta\phi$ is fixed at 1~mV.
Figure \ref{fig:result3}(c) displays $\Delta\tilde{v}_i$ for $i = 30, 50, 70, 100$.
The black dotted lines correspond to the transition points of the 2nd PC, $\phi_{\text{c-s}}$ and $\phi_{\text{s-cc}}$.
Note that the minimum points of $\Delta \tilde{v}_i$ around $\phi = 2 \text{--} 3$~mV shift to the larger $\phi$ as $i$ increases.
When $\Delta\tilde{v}_{50}$ reaches its minimum point before $\Delta\tilde{v}_{100}$, a significant gap emerges between the two rates of change.
As $\phi$ increases further, the gap between $\Delta \tilde{v}_{50}$ and $\Delta \tilde{v}_{100}$ starts to close around $\phi = 7$~mV.
Notably, the values of $\phi$ at which the gap starts to open and close are identical to the transition points of the 2nd PC.
This indicates that within the gapped region, $\Delta\tilde{v}_{\text{lar}}$ decreases more rapidly compared to $\Delta\tilde{v}_{\text{mid}}$, leading to an increase in the 2nd PC.
In contrast, in the gapless region for $\phi > 7$~mV, the 2nd PC transitions to a decrease.

\begin{figure}
 \includegraphics[width=\linewidth]{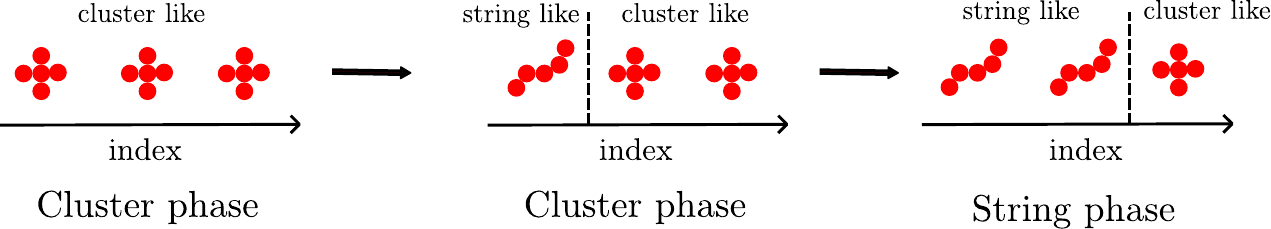}
  \caption{Illustration of asynchronous transitions. As the Coulomb interaction strengthens in the Cluster phase, transitions from cluster- to string-like local structures occur sequentially from low- to high-density regions within the system.}
  \label{fig:result4}
\end{figure}

Finally, we consider how the gap in the change rate of $v_i$ is related to the characteristics of each phase.
As shown in Fig.~\ref{fig:result3}(c), the rate $\Delta \tilde{v}_i$ evolves with $\phi$ through four stages: decreasing $\rightarrow$ increasing $\rightarrow$ decreasing $\rightarrow$ increasing.
We assume that this behavior corresponds to transitions in the local structure represented by $v_i$: Cluster phase $\rightarrow$ String phase $\rightarrow$ Coulomb Cluster phase $\rightarrow$ Coulomb String phase.
The shift in transition points of $\Delta \tilde{v}_i$ for different $i$ indicates that the transitions in local structures progress from low- to high-density regions as $\phi$ increases. [see Fig.~\ref{fig:result4}].
The opening of the gap, where $\Delta\tilde{v}_{50}$ transitions to an increase, can be interpreted as a sign that the majority of local structures have switched to the String phase.
Similarly, the closing of the gap, where $\Delta\tilde{v}_{50}$ transitions to a gradual decrease, indicates that most of the local structures have switched to the Coulomb Cluster phase.
In conclusion, the asynchronous transitions across local structures of varying densities lead to the gap in the change rate of $v_i$ and the non-monotonic behavior of the 2nd PC.
Note that, since the 2nd PC is a weighted sum of $v_i$, it effectively captures the dominant local structures.

%%%%%%%%%%%%%%%%%%%%%%%%%%%%%%%%%%%%
%%%%%%%%%%%%%%%%%%%%%%%%%%%%%%%%%%%%

\section{conclusion}
\label{sec:conclusion}

In this study, we proposed a novel method for classifying nonequilibrium disordered phases without prior knowledge.
The method involves representing a configuration of the system as a high-dimensional vector of local structural descriptors and applying PCA.
For a two-dimensional oppositely charged colloidal system, we demonstrated that the 2nd PC exhibits non-monotonic behavior with respect to interaction parameters and its turning points coincide with phase boundaries identified through radial distribution functions.
Furthermore, we also found that the weights of the 2nd PC have opposite signs for low- and high-density regions, and that transitions in local structures propagate sequentially from low- to high-density regions as the interaction parameters change.
The combination of these leads to the non-monotonic behavior of the 2nd PC.

These results demonstrate that our method can contribute to the study of transitions between globally disordered phases, such as liquid-liquid phase transitions \cite{Katayama2000, Dharma2020}.
Unlike previous studies \cite{Donkor2024,Boattini2019,Jung2025}, we treat the entire particle configuration as a single data point, rather than using descriptor vectors that encode the local environment around each particle.
This approach enables us to analyze the collective behavior of particles, which is essential for determining the phase of the system.
Furthermore, we employ PCA, which allows us to directly identify the features that most significantly influence the reduced representation.
This stands in contrast to neural-network-based approaches, which often rely on complex nonlinear mappings and may lack transparency.
This global and interpretable approach provides an important and complementary perspective to existing studies.

There are several issues for future research to improve and extend our method.
One limitation is that the dimensionality of the feature vectors used for PCA scales proportionally with the system size, which makes it challenging to apply this method to larger systems.
This issue could be mitigated by dividing configurations of the entire system into smaller, equally sized subsystems.
Another important issue is the dependence of the results on the choice of the local structural descriptor, defined by Eq.~\eqref{V} in this study.
While our findings are expected to be insensitive to the details of the descriptor, we observed that the non-monotonic behavior of the 2nd PC disappears when the cutoff scale $\xi$ in Eq.~\eqref{V} is significantly larger than the colloidal diameter $\sigma$.
This highlights the need for a model-free approach to optimize hyperparameters in the local structural descriptors.
Addressing these issues would enable our method to be applied to larger and more complex systems.

\acknowledgments
We thank Noriko Oikawa for her help on developing the oppositely charged colloidal model.
This work was supported by JST SPRING, Grant No. JPMJSP2139 and JSPS KAKENHI, Grant No. JP22K13983.

\appendix
\section{Derivation of the Coulomb Interaction in Our Model}
\label{sec:appendix_A}

In the following, we derive the screened Coulomb interaction term of Eq.~\eqref{u_el}.
This term represents the interaction between charges distributed on the surfaces of two colloids.
We assume that the ions in the electrolyte provide sufficient screening and the interaction range is short compared to the size of colloids.
Then, the interaction $U_{\text{el}}(r)$ between colloids separated by a center-to-center distance $r$ can be approximated using the interaction between two parallel charged walls as
\begin{equation}
    U_{\text{el}}(r)=\pi\frac{\sigma}{2}\int_{r-\sigma}^\infty{}dh\text{ }u_{\text{el}}(h),
    \label{U_r}
\end{equation}
where $u_{\text{el}}(h)$ is the electric potential energy per unit area of two walls spaced $h$, and $\sigma$ is the diameter of colloids.
This approximation is referred to as the Derjaguin approximation \cite{Israelachvili2011}.
To determine the form of $u_{\text{el}}(h)$, we calculate the electrostatic potential $\phi(x)$ for two parallel walls located at $x = \pm{}h/2$.
When $|q_{\alpha}\phi(x)|$, where $q_\alpha$ is the magnitude of the charge of the ion $\alpha$, is much less than $k_BT$ for any $\alpha$, the Poisson-Boltzmann equation within the region $-h/2\le{}x\le{}h/2$ is given by
\begin{equation}
    \frac{d^2\phi(x)}{dx^2}=\frac{1}{\lambda^2}\phi(x),
\end{equation}
where the Debye screening length $\lambda$ is described as
\begin{equation}
  \lambda=\sqrt{\frac{\epsilon{}k_{B}T}{\sum_\alpha{n_\alpha}q_\alpha^2}},
\end{equation}
with the bulk concentration $n_\alpha$ of the ion $\alpha$.
When the interacting colloids have the same surface charge $Q$, this equation is solved as 
\begin{equation}
    \phi(x)=\frac{{\lambda}Q}{\epsilon}\frac{\cosh(x/\lambda)}{\sinh(h/{2\lambda})}.
\end{equation}
By ignoring the constant terms, the potential can be approximated as
\begin{equation}
    u_{\text{el}}(h)=\frac{2\lambda{}Q^2}{\epsilon}\exp(-h/\lambda).
\end{equation}
For colloids with opposite surface charges, a similar calculation can be performed.
Finally, by denoting the surface charge of one colloid as $Q_1$ and the other as $Q_2$,
and substituting the calculated $u_{\text{el}}(h)$ into the Eq.~\eqref{U_r}, we obtain
\begin{equation}
    U_{\text{el}}(r)=\frac{\pi\lambda^2\sigma{}Q_1Q_2}{\epsilon}\exp\left(-\frac{r-\sigma}{\lambda}\right).
    \label{final_U_r}
\end{equation}

Next, we derive the relationship between the surface charge $Q_1$ and the surface potential $\phi_1$.
When the charge is uniformly distributed on the surface of the colloid, the potential $\phi(r)$ at a distance $r$ from the center of the colloid, for $r\ge\sigma/2$, follows
\begin{equation}
    \frac{d^2\phi(r)}{dr^2}+\frac{2}{r}\frac{d\phi(r)}{dr}=\frac{1}{\lambda^2}\phi(r).
\end{equation}
This equation is solved as
\begin{equation}
    \phi(r)=\frac{\sigma\phi_1}{2r}\exp\left(-\frac{2r-\sigma}{2\lambda}\right).
\end{equation}
By the relationship between the electric field and the surface charge, we can obtain
\begin{equation}
    Q_1=-\epsilon\left.\frac{d\phi}{dr}\right|_{r=\frac{\sigma}{2}}=\frac{\epsilon}{\lambda}\left(1+\frac{2\lambda}{\sigma}\right)\phi_1.
\end{equation}
By substituting this expression into Eq.~\eqref{final_U_r}, we get the Coulomb potential of Eq.~\eqref{u_el}.

\section{Sampling of Nonequilibrium States and System Dynamics}
\label{sec:appendix_B}

%%%%%%%%%%%%%%%%%%%%%%%%%%%%%%%%%%%%
\begin{figure}[t]
 \includegraphics[width=\linewidth]{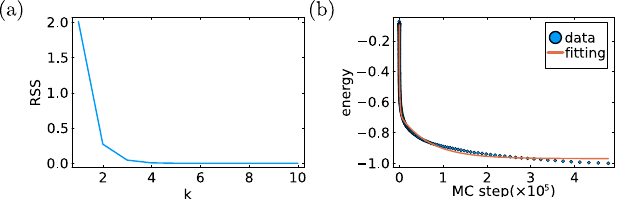}
  \caption{Panel (a) plots along $k$ the residual sum of squares (RSS) between the normalized energy $E(t)/|E(5\times10^5)|$ and fitted function $f_k(t)$ for the Debye length $\lambda=2.5$~nm and the surface potential $\phi=1$~mV. Panel (b) represents $E(t)/|E(5\times10^5)|$ and the fitted triple exponential function $f_3(t)$.}
  \label{fig:appendix1}
\end{figure}
%%%%%%%%%%%%%%%%%%%%%%%%%%%%%%%%%%%%

In our Monte Carlo simulations, colloids form locally favored structures, and their dynamics become frozen before reaching equilibrium from an initial random configuration.
Since our objective is to sample these transient structures, it is necessary to determine the timescale over which the system completes its initial structural relaxation from a random configuration to locally preferred arrangements.
To this end, we fit the time-dependent potential energy with a multi-exponential function to extract characteristic timescales associated with both fast and slow relaxation processes.
Based on these timescales, we determine the sampling time corresponding to the completion of the fast structural relaxation.

Specifically, we use the following fitting function $f_k(t)$ to model the time evolution of the potential energy:
\begin{equation}
f_k(t)=\sum_{i=1}^{k}{a_i{}\exp\left(-\frac{t}{\tau_i}\right)}+c,
\end{equation}
where $a_i$ is the coefficient of the $i$-th exponential function, $\tau_i$ is its corresponding relaxation time (with $\tau_i<\tau_{i+1}$), and $c$ represents the long-time asymptotic value.
We apply this fitting to the normalized potential energy $E(t)/|E(5\times10^5)|$ over the range from $t=0$ to $5\times10^5$ MC steps.
Here, one MC step consists of $N_\text{tot}$ displacement attempts, and $N_\text{tot}$ is the total particle number.
To determine the optimal number of exponential terms $k$, we calculate the residual sum of squares (RSS) between data and $f_k(t)$ for varying $k$.
Figure~\ref{fig:appendix1}~(a) shows the RSS for the Debye length $\lambda = 2.5 \ \text{nm}$ and the surface potential $\phi = 1 \ \text{mV}$.
The RSS significantly decreases up to $k=3$, beyond which the improvement saturates.
Accordingly, we adopt $k=3$ and show the resulting triple exponential fitting $f_3(t)$ in Fig.~\ref{fig:appendix1}~(b), which yields characteristic timescales $\tau_1=69$, $\tau_2=1565$, and $\tau_3=68030$ (in units of MC steps).
The characteristic time $\tau_3$, which differs from others by more than one order of magnitude, corresponds to the slowest dynamics toward the globally stable state, while $\tau_2$ characterizes the faster relaxation process falling into locally favored structures.
After we conduct the same protocol over the examined parameter space, we find that the optimal number $k$ is also 3, and $\tau_2$ doesn't change significantly, and its average value $\tilde{\tau}_2=1300$.
Therefore, we sample particle configurations at $3\tilde{\tau}_2=3900$ MC steps to ensure that the initial relaxation is complete.

%%%%%%%%%%%%%%%%%%%%%%%%%%%%%%%%%%%%
\begin{figure}
 \includegraphics[width=\linewidth]{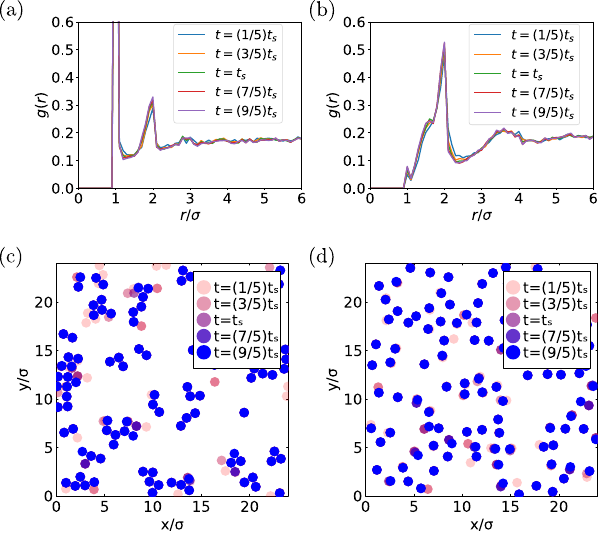}
  \caption{Panels~(a) and (b) plot the RDFs for positively charged colloids $g_{pp}$ from $t=(1/5)t_s$ to $(9/5)t_s$ in increments of $(2/5)t_s$. Panel (a) corresponds to ($\phi$, $\lambda$)=(1mV, 2.5nm), and panel (b) to (20mV, 50nm). The RDFs remain nearly unchanged around the sampling time $t_s$. Panel~(c) shows the coordinates of positively charged colloids for ($\phi$, $\lambda$)=(1mV, 2.5nm) at different times, and panel~(d) represents those for (20mV, 50nm). As time increases, the color and transparency of markers change from thin red to thick blue. It is evident that the positions of the colloids remain essentially fixed beyond $t_s$.}
  \label{fig:appendix2}
\end{figure}
%%%%%%%%%%%%%%%%%%%%%%%%%%%%%%%%%%%%

Next, we show that the structure changes very little around the sampling time $t_s=3900$.
Figure~\ref{fig:appendix2}(a) and (b) show the RDFs from $t=(1/5)t_s$ to $(9/5)t_s$ in increments of $(2/5)t_s$.
We observe that the RDFs remain nearly unchanged after $t_s$. 
Furthermore, Fig.~\ref{fig:appendix2}(c) and (d) display the coordinates of the positively charged colloids at several times, with particle colors varying from thin red (earlier times) to thick blue (later times).
The overlapping of blue-colored particles with earlier ones clearly indicates that the particles become dynamically frozen after $t_s$.
The figures demonstrate that the system is trapped in a long-lived metastable state.

\section{Applying Simulated Annealing}
\label{sec:appendix_C}

Once the slowest relaxation is complete and thermal equilibrium is reached, the final configurations can differ significantly from the nonequilibrium phases shown in Fig.~\ref{fig:phase_morphology}.
In equilibrium, the colloids are expected to form either a phase-separated crystal or an ionic crystal, depending on the competition between van der Waals and Coulomb interactions.
To investigate the colloidal configurations near equilibrium, we use simulated annealing, a technique in which the temperature is gradually reduced from a value higher than the target temperature.
This technique allows for effective sampling of equilibrium states of systems that exhibit slow relaxation.

%%%%%%%%%%%%%%%%%%%%%%%%%%%%%%%%%%%%
\begin{figure}
 \includegraphics[width=\linewidth]{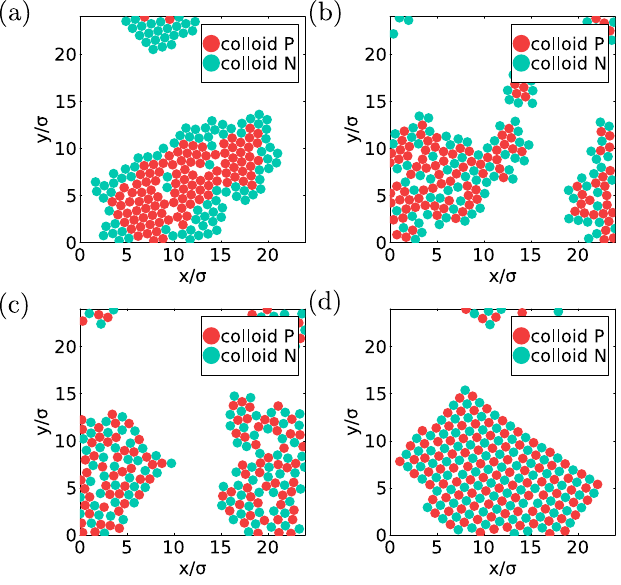}
  \caption{The snapshot sampled by the simulated annealing. Panels (a)-(d) are the configurations corresponding to the Cluster, String, Coulomb Cluster, and Coulomb String phases, respectively.}
  \label{fig:appendix3}
\end{figure}
%%%%%%%%%%%%%%%%%%%%%%%%%%%%%%%%%%%%

In the following simulations, the initial temperature is set to 20 times higher than 300 K and gradually reduced to 300 K.
The results are shown in Fig.~\ref{fig:appendix3}.
For the Cluster and String phases shown in Figs.~\ref{fig:appendix3}(a) and~\ref{fig:appendix3}(b), we observe well-developed cluster- and string-like structures.
In the Coulomb Cluster phase shown in Fig.~\ref{fig:appendix3}(c), a mixture of short string-like and ionic crystal-like structures is found.
For the Coulomb String phase shown in Fig.~\ref{fig:appendix3}(d), we can see that the ionic crystal.
While the results in Figs.~\ref{fig:appendix3}(a) and \ref{fig:appendix3}(d) align with expectations, Fig.~\ref{fig:appendix3}(b) and (c) display an irregular structure that deviates from typical phase-separated crystals and ionic crystals. This irregularity occurs because the parameters in Fig.~\ref{fig:appendix3}(b) and (c) lie near the boundary between ionic and non-ionic crystal regimes, leading to extremely slow relaxation.

\bibliographystyle{apsrev4-2}
\bibliography{paper.bib}
\end{document}